\documentclass{appolb}
\usepackage{amsfonts}
\usepackage{graphicx}
\usepackage{bm}
\usepackage{slashed}
\usepackage{comment}
\usepackage{pstricks}
\usepackage{color}
\usepackage{xcolor}
\usepackage{amsmath}
\usepackage{amssymb}
\usepackage{braket}
\usepackage{multirow}
\usepackage{hyperref}
\usepackage{stmaryrd}
\usepackage{cite}

\makeatletter
\DeclareRobustCommand{\cev}[1]{%
  {\mathpalette\do@cev{#1}}%
}
\newcommand{\do@cev}[2]{%
  \vbox{\offinterlineskip
    \sbox\z@{$\m@th#1 x$}%
    \ialign{##\cr
      \hidewidth\reflectbox{$\m@th#1\vec{}\mkern4mu$}\hidewidth\cr
      \noalign{\kern-\ht\z@}
      $\m@th#1#2$\cr
    }%
  }%
}
\makeatother

\begin{document}
\title{Small-$x$ Quark and Gluon Helicity Contributions to the Proton Spin Puzzle
\thanks{Presented at XXIX Cracow Epiphany Conference on Physics at the Electron-Ion Collider and Future Facilities, Cracow, Poland, January 16-19, 2023.}%
}
\author{Yossathorn Tawabutr
\address{
Department of Physics, University of Jyv\"askyl\"a \\ P.O. Box 35, 40014 University of Jyv\"askyl\"a, Finland
}
\address{
Helsinki Institute of Physics \\ P.O. Box 64, 00014 University of Helsinki, Finland
}
}
\maketitle
\begin{abstract}
A small-$x$ helicity evolution has been derived in 2016-18 and received an important modification in 2022. This article discusses its general framework and summarizes the recent theoretical developments, including the asymptotic behaviors of helicity PDFs and $g_1$ structure function at small $x$. The latest fits to various polarized scattering data are also discussed. The results from this research program will provide important theoretical inputs for the future polarized small-$x$ measurements at the electron-ion collider (EIC).
\end{abstract}
  
\section{Introduction}

The proton spin puzzle is a longstanding problem in quantum chromodynamics: how much of the proton spin is distributed among the quarks and gluons? In the late 1980s, the European Muon Collaboration published the results \cite{EMC1, EMC2} that only 14\% of proton's helicity comes from that of the constituent quarks, contradicting the popular belief at the time. Later experimental results \cite{RHIC_spin1, RHIC_spin2} with significant improvement in the uncertainty still confirm that the quark's helicity alone does not add up to the proton's helicity of $\frac{1}{2}$. 

Various spin sum rules \cite{Leader:2013jra} are proposed as more general descriptions for how the proton's spin is distributed within the partons inside. In particular, we focus on the Jaffe-Manohar sum rule \cite{JM} that allows us to separate the proton's helicity into the contributions from the helicities, $S_q$ and $S_g$, and the orbital angular momenta, $L_q$ and $L_g$, of quarks and gluons inside. Explicitly, we have
\begin{align}\label{JMsumrule}
\frac{1}{2} &= S_q + S_g + L_q + L_g\,.
\end{align}
In particular, the quark's and gluon's helicity contributions can be written in terms of the respective helicity distribution functions (hPDF) as
\begin{subequations}\label{StohPDF}
\begin{align}
S_q(Q^2) &= \frac{1}{2}\int_0^1dx\,\Delta\Sigma(x,Q^2)\,, \label{StohPDFqk} \\
S_g(Q^2) &= \int_0^1dx\,\Delta G(x,Q^2)\,, \label{StohPDFgl}
\end{align}
\end{subequations}
where $\Delta G$ is the gluon hPDF, that is, the difference between the PDFs of gluons with positive and negative helicity:
\begin{align}\label{DeltaGdiff}
\Delta G(x,Q^2) &= g^+(x,Q^2) - g^-(x,Q^2) \, .
\end{align}
In Eq. \eqref{StohPDFqk}, $\Delta\Sigma$ is the flavor-singlet quark hPDF, which can be written out as
\begin{align}\label{DeltaSigmadecomp}
\Delta\Sigma(x,Q^2) &= \sum_f\left[\Delta f(x,Q^2) + \Delta\bar{f}(x,Q^2)\right]  .
\end{align}
Here, $\Delta f$ and $\Delta\bar{f}$ are hPDFs of the quark and the antiquark of flavor $f$, respectively. They are defined similarly to Eq. \eqref{DeltaGdiff}. Note that the summation in Eq. \eqref{DeltaSigmadecomp} is only over the light quark flavors, typically up, down and strange.

Owing to the integrals over Bjorken $x$ in Eqs. \eqref{StohPDF} going all the way down to zero, no experimental measurement can completely determine the quark and gluon helicity contributions to the sum rule. The goal of our spin program is to fill this gap by constructing the small-$x$ evolution for hPDFs, akin to the BK/JIMWLK evolution for unpolarized PDFs, with the ultimate goal of determining their small-$x$ asymptotics. The results will complement experimental measurements, including those of the future electron-ion collider (EIC) \cite{Accardi:2012qut}, to provide a more complete understanding of the proton spin puzzle.

The first version of small-$x$ helicity evolution is the Kovchegov-Sievert-Pitonyak (KPS) equation, which evolves the quark \cite{Kovchegov:2015zha, Kovchegov:2015pbl, Kovchegov:2018znm} and the gluon \cite{Kovchegov:2017lsr} hPDFs separately, with the solutions provided in \cite{Kovchegov:2016weo, Kovchegov:2017jxc, Kovchegov:2020hgb} and a successful preliminary global fit performed in \cite{Adamiak:2021ppq}. The equation resums $\alpha_s\ln^2(1/x)$, where $\alpha_s$ is the strong coupling constant. Its corrections at single-logarithmic approximation (SLA) are calculated in \cite{Kovchegov:2021lvz}, with a summary given in \cite{Tawabutr:2021xrr}. 

More recently, an additional contribution to the quark hPDF has been discovered \cite{Cougoulic:2022gbk}, leading to the coupling between the quark and gluon hPDFs in their small-$x$ evolution. This leads to the new KPS-Cougoulic-Tarasov-Tawabutr (KPSCTT) evolution, which describes both the quark and gluon hPDFs at small $x$ to the double-logarithmic approximation (DLA) \cite{Cougoulic:2022gbk}. The asymptotic solutions in various limits are given in \cite{newNcNf, JeremyLargeNc}. Furthermore, the complete SLA corrections are in progress \cite{SLAops}. The framework can also be extended to study the orbital angular momentum (OAM) at small $x$ \cite{Kovchegov:2019rrz, BrandonOAM}. 

This article provides a brief overview of recent developments around the KPSCTT evolution equation for the quark and gluon hPDFs at small $x$. Section \ref{sect:evol} defines the necessary objects in the dipole picture of the helicity-dependent deep-inelastic scattering (DIS) process and outlines how the small-$x$ evolution is constructed in terms of them. Then, Sections \ref{sect:Nc} and \ref{sect:NcNf} discuss the evolution equations and the surrounding development in the limit of large $N_c$ \cite{tHooft:1973alw} and large $N_c\& N_f$ \cite{Veneziano:1976wm}, respectively, with $N_c$($N_f$) the number of quark colors(light flavors). Section \ref{sect:pheno} discusses the recent phenomenological developments. Finally, we conclude and discuss future prospects of the program in Section \ref{sect:conclusion}.

\section{Polarized Dipole Picture and The Small-$x$ Evolution}\label{sect:evol}

The deep-inelastic scattering (DIS), in which an electron exchanges a virtual photon with a hadron target in such a way that the target is broken apart, is a useful process to probe the PDFs of partons inside the target \cite{Peskin, Yuribook}. At small $x$, the process can be described to a good approximation by a convolution between the $e^-\to q\bar{q}$, i.e. the dipole, wave function and the cross section of the dipole-target scattering process \cite{Dipole_Bertsch, Dipole_Kopeliovich, Dipole_Mueller, Dipole_Nikolaev}. This results in the overall cross section proportional to the CGC average \cite{Yuribook} over the target's wave function of the trace of a Wilson line and its adjoint,
\begin{align}\label{S10}
S(x^2_{10}, zs) &= \frac{1}{N_c} \int d^2\left(\frac{\underline{x}_0+\underline{x}_1}{2}\right) \left\langle \text{tr}\left[ V_{\underline{0}}V_{\underline{1}}^{\dagger} \right] \right\rangle (z) \, .
\end{align}
Here, $V_{\underline{i}} \equiv V_{\underline{x}_i}[\infty,-\infty]$ is the light-cone Wilson line at transverse position, $\underline{x}_i$, which is defined generally as 
\begin{align}\label{WilsonLine}
V_{\underline{i}}[x_f^-,x_i^-] &= \mathcal{P}\exp\left[ ig\int_{x_i^-}^{x_f^-}dx^-A^+(x^+=0, x^-, \underline{x}_i)  \right] ,
\end{align}
where $\mathcal{P}$ is the path-ordering operator. Diagrammatically, the Wilson line corresponds to multiple $t$-channel exchanges of longitudinal gluons at the eikonal level, i.e. the leading power of the squared center-of-mass energy, $s$.

For hPDFs, the story is similar. However, we now require a helicity-dependent DIS, for which the cross section of the process where the electron and the target are aligned in helicities is subtracted by that of the process where the helicities are anti-aligned. The analogous dipole picture holds, but with either the quark or the antiquark in the $q\bar{q}$ dipole now becoming ``polarized'', that is, the cross section becomes proportional to the product between its helicity and the target's helicity. The polarized (anti)quark now corresponds to a sub-eikonal (suppressed by a power of $s$) ``polarized Wilson line''. The latter can be written as \cite{Kovchegov:2018znm, Kovchegov:2017lsr, Cougoulic:2022gbk} (notice the helicity structure)
\begin{align}\label{Vpol}
V_{\underline{1}', \underline{1};\,\sigma',\sigma}\big|_{\text{sub-eikonal}} &= \sigma\,\delta_{\sigma\sigma'}\,\delta^2(\underline{x}_{1'}-\underline{x}_1)\,V_{\underline{1}}^{\text{pol}[1]} + \delta_{\sigma\sigma'}\,V_{\underline{1}', \underline{1}}^{\text{pol}[2]}  ,
\end{align}
where $V_{\underline{1}}^{\text{pol}[1]}$ is the ``type-1 polarized Wilson line'' that describes sub-eikonal quark ($V_{\underline{1}}^{\text{q}[1]}$) and gluon ($V_{\underline{1}}^{\text{G}[1]}$) exchanges antisymmetric in helicity:
\begin{align}\label{V1pol}
V_{\underline{1}}^{\text{pol}[1]} &= V_{\underline{1}}^{\text{q}[1]} + V_{\underline{1}}^{\text{G}[1]}
\end{align}
with
\begin{subequations}\label{V1polqg}
\begin{align}
V^{\text{q[1]}}_{\underline{1}} &=   \frac{g^2P^+}{2s} \, \int_{-\infty}^{\infty}dx_1^-\int_{x_1^-}^{\infty}dx^-_2\, V_{\underline{1}}[\infty,x_2^-]\,t^b \left[\psi(x_2^-,\underline{x}_1)\right]_{\beta}    \label{Vq1} \\
&\;\;\;\;\;\times U^{ba}_{\underline{1}}[x_2^-,x_1^-] \left[\gamma^+\gamma_5\right]_{\alpha\beta}  \left[\bar{\psi}(x_1^-,\underline{x}_1)\right]_{\alpha}t^a\,V_{\underline{1}}[x_1^-,-\infty] \, , \notag  \\ 
V^{\text{G[1]}}_{\underline{1}} &=  \frac{igP^+}{s} \int_{-\infty}^{\infty}dx^-\,V_{\underline{1}}[\infty,x^-] \, F^{12}(x^-,\underline{x}_1) \,V_{\underline{1}}[x^-,-\infty]\,.  \label{VG1} 
\end{align}
\end{subequations}
Here, $P^+$ is the (large) longitudinal momentum of the target. These type-1 polarized Wilson lines induce the definition of the ``type-1 polarized dipole amplitude'',
\begin{align}\label{Q10}
Q(x^2_{10}, zs) &= \frac{zs}{2N_c} \int d^2\left(\frac{\underline{x}_0+\underline{x}_1}{2}\right)  \\
&\;\;\;\;\;\times \text{Re} \left\langle \text{T}\,\text{tr}\left[V_{\underline{1}}^{\text{pol}[1]}V_{\underline{0}}^{\dagger} \right] + \text{T}\,\text{tr}\left[ V_{\underline{0}} V_{\underline{1}}^{\text{pol}[1]\dagger} \right] \right\rangle (zs) \, , \notag
\end{align}
where T is the light-cone time-ordering operator. 

In Eq. \eqref{Vpol}, $V_{\underline{1}', \underline{1}}^{\text{pol}[2]}$ is the ``type-2 polarized Wilson line'' that describes sub-eikonal gluon exchange symmetric in helicity\footnote{It is curious how a sub-eikonal contribution that comes in with the identity matrix in helicity space contributes to helicity in general. This may result from the difference between the helicity basis employed in the Brodsky-Lepage spinor \cite{LCPT1, LCPT2} and the true helicity basis that is based on the direction of the particle's momentum. See footnote 3 in \cite{Cougoulic:2022gbk} for more details.}. Its explicit form is
\begin{align}
V^{\text{pol[2]}}_{\underline{1}',\,\underline{1}} = V^{\text{G[2]}}_{\underline{1}',\,\underline{1}} &= - \frac{iP^+}{s} \int_{-\infty}^{\infty}dz^-\,d^2\underline{z}\,V_{\underline{1}'}[\infty,z^-]\,\delta^2(\underline{x}_{1'}-\underline{z})    \label{VG2} \\
&\;\;\;\;\;\times \cev{\underline{D}}^i(z^-,\underline{z})\,\vec{\underline{D}}^i (z^-,\underline{z}) \, \delta^2(\underline{x}_1-\underline{z}) \, V_{\underline{1}}[z^-,-\infty]   \, , \notag
\end{align}
where the covariant derivatives are defined such that $\vec{\underline{D}}^i = \vec{\underline{\partial}}^i - ig\underline{A}^i$ and $\cev{\underline{D}}^i = \cev{\underline{\partial}}^i + ig\underline{A}^i$. This type of the Wilson line was not included in the original KPS equation \cite{Kovchegov:2015pbl, Kovchegov:2018znm}, but it was discovered to add a vital contribution to helicity at small $x$, leading to a more recent KPSCTT equation \cite{Cougoulic:2022gbk}. Now, the transverse integral of 
\begin{align}\label{trV2}
&\text{Re} \left\langle \text{T}\,\text{tr}\left[V_{\underline{1}',\underline{1}}^{\text{pol}[2]}V_{\underline{0}}^{\dagger} \right] + (\text{c.c.}) \right\rangle (zs)
\end{align}
can be more conveniently written as a transverse integral (with a different kernel) of 
\begin{align}\label{G2}
G_2(x^2_{10},zs) &= \frac{zs}{2N_c} \int d^2\left(\frac{\underline{x}_0+\underline{x}_1}{2}\right) \frac{\epsilon^{ij}\underline{x}_{10}^j}{x^2_{10}}  \\
&\;\;\;\;\;\times  \left\langle \text{tr}\left[V_{\underline{0}}^{\dagger}V_{\underline{1}}^{i\,\text{G}[2]} \right] +  \text{tr}\left[ V_{\underline{1}}^{i\,\text{G}[2]\dagger}V_{\underline{0}}  \right] \right\rangle (zs) \, , \notag
\end{align}
where $\underline{x}_{10} = \underline{x}_1-\underline{x}_0$ and
\begin{align}\label{ViG2}
V_{\underline{1}}^{i\,\text{G}[2]} &= \frac{P^+}{2s}\int_{-\infty}^{\infty}dx^-\,V_{\underline{1}}[\infty,x^-]\left(\vec{D}^i(x^-,\underline{x}_1)-\cev{D}^i(x^-,\underline{x}_1)\right) \\
&\;\;\;\;\;\times  V_{\underline{1}}[x^-,-\infty] \, . \notag
\end{align}
We define the quantity in Eq. \eqref{G2} as the ``type-2 polarized dipole amplitude'' \cite{Cougoulic:2022gbk}.

At small $x$, the polarized dipole amplitudes allow us to calculate the $g_1$ structure function and parton hPDFs. Explicitly, we have
\begin{subequations}\label{g1hPDFs}
\begin{align}
g_1(x,\,Q^2)  &= - \frac{N_c}{4\pi^3}\sum_f Z_f^2 \int\limits_{\Lambda^2/s}^1\frac{dz}{z}  \int \frac{dx^2_{10}}{x^2_{10}} \label{g1} \\
&\;\;\;\;\;\times \left[ Q(x^2_{10},\,zs) + 2G_2(x^2_{10},\,zs)   \right] , \notag \\
\Delta\Sigma(x,\,Q^2) &=  - \frac{N_cN_f}{2\pi^3} \int_{\Lambda^2/s}^1 \frac{dz}{z} \int \frac{dx^2_{10}}{x_{10}^2} \left[Q(x^2_{10},zs) + 2 G_2(x^2_{10},zs) \right] , \label{qkhPDF} \\  
\Delta G(x, Q^2) &= \frac{2N_c}{\alpha_s\pi^2} \left[1+x^2_{10}\frac{\partial}{\partial x^2_{10}} \right] G_2(x^2_{10},zs) \bigg|_{x^2_{10} = 1/Q^2} \,  . \label{glhPDF}
\end{align}
\end{subequations}
The small-$x$ evolution for helicity is more conveniently expressed in terms of the polarized dipole amplitudes, the results of which allow us to determine the $g_1$ structure function and the parton hPDFs using Eqs. \eqref{g1hPDFs}. 

The KPSCTT evolution is formulated in term of a polarized $q\bar{q}$ dipole interacting with the target in a shockwave\footnote{The notion follows from the fact that the interaction time scale is short at small $x$.}. Each step of evolution corresponds to shrinking the shockwave so that it excludes one extra parton emission and absorption, for which we now employ the ``light-cone operator treatment'' (LCOT) method to calculate \cite{Kovchegov:2018znm, Kovchegov:2017lsr, Cougoulic:2022gbk}. The LCOT method is a mix between the light-cone perturbation theory (LCPT) \cite{LCPT1, LCPT2} and the background field method \cite{Abbott:1980hw, Abbott:1981ke}. This gives an integral evolution equation that relates the original dipole to not only other dipoles, but also quadrupoles \cite{Kovchegov:2015pbl}, so that the complete solution for the dipole also requires us to solve the evolution equation for the quadrupole, and so on. This is the helicity counterpart of Balitsky's hierarchy \cite{Balitsky:1995ub, Balitsky:1998ya}, making it difficult to directly solve the evolution equation for the polarized dipole amplitudes.

\section{Large-$N_c$ Limit: The Gluonic World}\label{sect:Nc}

The KPSCTT evolution becomes a closed system of integral equations once we take the large-$N_c$ limit \cite{tHooft:1973alw}, make the mean-field approximation and remain in the regime of moderately small $x$ in which the unpolarized small-$x$ evolution has not taken off. Furthermore, the equation linearizes if we perform the calculation within the pre-saturation regime\footnote{Since the KPSCTT evolution is double-logarithmic, resumming $\alpha_s\ln^2(1/x)$, its ``small-$x$'' regime kicks in at larger $x=x_0$ than the small-$x$ regime for the unpolarized BK/JIMWLK small-$x$ evolution. In this regime of $0.01\lesssim x\lesssim 0.1$, one can employ the KPSCTT evolution while keeping the unpolarized dipole amplitude, $S(x^2_{10},zs)$, at its moderate-$x$ initial condition. For $x\lesssim 0.01$, we also need to include the SLA corrections to the KPSCTT evolution as well, c.f. \cite{Kovchegov:2021lvz} for further discussion.} \cite{Kovchegov:2015pbl, Cougoulic:2022gbk}.

As quark exchanges are suppressed at large $N_c$, the type-1 polarized dipole amplitude, $Q(x^2_{10},zs)$, is approximated by the term coming only from the type-1 sub-eikonal gluon exchange, denoted by $G(x^2_{10},zs)$. Then, the KPSCTT equation reduces to a system of linear integral equations that can be written schematically as\footnote{The complete evolution equation is given in \cite{Cougoulic:2022gbk}.}
\begin{align}\label{EvolEqNc}
&\begin{pmatrix}
G(x^2_{10},zs) \\ \Gamma(x^2_{10},x^2_{21},zs) \\ G_2(x^2_{10},zs) \\ \Gamma_2(x^2_{10},x^2_{21},zs)
\end{pmatrix} = 
\begin{pmatrix}
G^{(0)}(x^2_{10},zs) \\ \Gamma^{(0)}(x^2_{10},x^2_{21},zs) \\ G_2^{(0)}(x^2_{10},zs) \\ \Gamma_2^{(0)}(x^2_{10},x^2_{21},zs)
\end{pmatrix} \\
&\;\;\;+ 
\frac{\alpha_sN_c}{2\pi} \int \frac{dz'}{z'} \int \frac{dx^2_{32}}{x^2_{32}}\;K(x^2_{10},x^2_{21},zs;x^2_{32},z's)
\begin{pmatrix}
G(x^2_{32},z's) \\ \Gamma(x^2_{10},x^2_{32},z's) \\ G_2(x^2_{32},z's) \\ \Gamma_2(x^2_{10},x^2_{32},z's)
\end{pmatrix} , \notag
\end{align}
where the $4\times 4$ matrix, $K$, is the evolution kernel and $\Gamma$($\Gamma_2$) is the ``neighbor dipole amplitude'' for $G$($G_2$). In particular, $\Gamma(x^2_{10},x^2_{21},zs)$ is the type-1 amplitude with physical transverse size $x_{10}$ but lifetime $x^2_{21}z$ \cite{Kovchegov:2015pbl, Kovchegov:2018znm, Cougoulic:2022gbk}. The initial condition should ideally be taken based on the amplitudes at moderate $x$, although the evolution generates the power-law growth in the amplitudes that quickly dominates the effects of the initial condition \cite{Kovchegov:2016weo, Kovchegov:2017jxc, Cougoulic:2022gbk}.

Solving Eq. \eqref{EvolEqNc} and plugging the high-energy asymptotics of the amplitudes into Eqs. \eqref{g1hPDFs}, we obtain the small-$x$ asymptotics of the form
\begin{align}\label{asympNc}
g_1(x,\,Q^2)  &\sim \Delta\Sigma(x,\,Q^2) \sim \Delta G(x, Q^2) \sim \left(\frac{1}{x}\right)^{\alpha_h\sqrt{\alpha_sN_c/2\pi}}\,,
\end{align}
where the ``intercept'', $\alpha_h$, is a parameter associated with the large-$N_c$ evolution kernel. It is $\alpha_h$ that we need to determine when we solve the evolution equations of this kind for the small-$x$ asymptotics. 

In 1990s, Bartels et al. (BER) calculated the parton helicity using the infrared renormalization evolution equation (IREE) approach, whose results included the small-$x$ regime \cite{Bartels:1996wc}. In the large-$N_c$, i.e. pure-gluon, limit, the intercept was found to be $\alpha_h=3.66$.

Originally, with the KPS evolution, the quark and gluon hPDFs relate to different dipole amplitudes that evolve separately. This results in $\alpha_h=2.31$ for quarks \cite{Kovchegov:2016weo, Kovchegov:2017jxc} and $\alpha_h=1.88$ for gluons \cite{Kovchegov:2017lsr}, both of which significantly disagree with BER \cite{Bartels:1996wc}. Now, with the KPSCTT evolution, both hPDFs have the same intercept. A numerical calculation performed in \cite{Cougoulic:2022gbk} gives $\alpha_h=3.66$, seemingly agreeing with BER \cite{Bartels:1996wc}. However, the most recent calculation shows that the analytic expressions for $\alpha_h$ disagree, with $\alpha_h\simeq 3.661$ for KPSCTT and $\alpha_h\simeq 3.664$ for BER \cite{JeremyLargeNc}. Hence, in the end, there remains a slight discrepancy in the small-$x$ asymptotic behaviors predicted by KPSCTT and BER, which requires further study.

\section{Large-$N_c\& N_f$ Limit: Bringing Back the Quarks}\label{sect:NcNf}

Recall from Section \ref{sect:evol} that the polarized Wilson line involves both gluon and quark exchanges, in contrast to the unpolarized Wilson line that only involves gluon exchanges. As a result, the large-$N_c$ limit, in which all quark exchanges are neglected in favor of the gluons \cite{tHooft:1973alw}, is not as good of an approximation in the helicity evolution as it has been in the unpolarized small-$x$ evolution like BFKL or BK \cite{Kovchegov:2015pbl}. This warrants a more realistic limit in which the KPSCTT evolution becomes closed: the large-$N_c\& N_f$ limit \cite{Veneziano:1976wm}. Here, we take $N_f/N_c$ to be a fixed parameter of order unity, as opposed to the large-$N_c$ limit in which we assumed $N_f\ll N_c$ right away. Consequently, the quark exchanges are now included in the calculation, although the gluon lines are still approximated by color-octet $q\bar{q}$ pairs. This requires us to distinguish between a type-1 dipole, $Q(x^2_{10},zs)$, that contains a true quark and another type-1 dipole, ${\widetilde G}(x^2_{10},zs)$, whose $q$ and $\bar{q}$ are both parts of gluons \cite{Kovchegov:2015pbl, Kovchegov:2018znm, Cougoulic:2022gbk}. The type-2 dipole is still described by $G_2(x^2_{10},zs)$.

The large-$N_c\& N_f$ evolution equation is a closed system of integral equations, which linearize under the mean-field approximation. The system involves 3 dipole amplitdes -- $Q$, ${\widetilde G}$ and $G_2$ -- together with their respective neighbor dipole amplitudes -- $\bar{\Gamma}$, ${\widetilde \Gamma}$ and $\Gamma_2$. The equation can be written schematically in a similar manner as Eq. \eqref{EvolEqNc} 
but with 6 distinct polarized dipole amplitudes in the row vectors and a different kernel made of a $6\times 6$ matrix.

With $N_c=3$ quark colors, as long as we keep the number of flavors, $N_f$, at 5 or below, the resulting small-$x$ asymptotics for the $g_1$ structure function and the hPDFs obey the power law with intercept $\alpha_h^{(N_f)}$ similar to the large-$N_c$ case, such that for each $N_f$, we have \cite{newNcNf}
\begin{align}\label{asympNcNf}
g_1(x,\,Q^2)  &\sim \Delta\Sigma(x,\,Q^2) \sim \Delta G(x, Q^2) \sim \left(\frac{1}{x}\right)^{\alpha_h^{(N_f)}\sqrt{\alpha_sN_c/2\pi}}\,.
\end{align}
The numerical calculation shows that $\alpha_h^{(N_f)}$ decreases as $N_f$ increases \cite{newNcNf}. However, $\alpha_h^{(N_f)}$ appears to differ from the BER intercepts \cite{Bartels:1996wc} by up to $3\%$ for $N_f=2,3$ and $4$. This is a much greater discrepancy than what we got at large $N_c$, and it provides an even clearer evidence that the small-$x$ asymptotic behaviors obtained from the KPSCTT and BER evolutions are different. A possible cause could be a minor difference in the ways hPDFs and $g_1$ are defined within the IREE and the CGC frameworks. Although the resolution of this issue remains under investigation, the difference between the current and the final results should not exceed the present level of discrepancy, which is relatively minor.

Another remarkable result from the large-$N_c\& N_f$ evolution is in the cases where $N_f\geq 6$. Here, the hPDFs and the $g_1$ structure function not only grow in the magnitudes as a power law with $\frac{1}{x}$, but the asymptotic form also gets multiplied by a sinusoidal function of $\frac{1}{x}$ \cite{newNcNf}. However, since the derivation of KPSCTT is performed using the massless approximation, the equation becomes potentially inaccurate when 6 quark flavors are included.

Finally, it is worth noting that the small-$x$ asymptotic behavior given in Eqs. \eqref{asympNc} and \eqref{asympNcNf} are only valid at the DLA level of the helicity evolution. For sufficiently small $x$, the SLA corrections and the unpolarized BK evolution of $S(x^2_{10},zs)$ has to be included in the calculation of the asymptotic form. In such regimes, the running of the coupling should also come into consideration. Recently, all such corrections for the original KPS equation have been derived \cite{Kovchegov:2021lvz}. Now, with the additional dipole amplitude of type 2, the corrections are expected to receive more contributions. This is a work in progress \cite{SLAops}. Qualitatively, since the BK evolution leads to parton saturation, it is reasonable to expect that the growth in Eqs. \eqref{asympNc} and \eqref{asympNcNf} will be tamed at smaller $x$ with the complete SLA corrections. As for the case of $N_f=6$, the results show that the period of oscillation is long, spanning many units of rapidity. Thus, it is also likely that the SLA corrections will come in and change the asymptotic form in a qualitative way before any sign flip could take place.

\section{Phenomenological Development}\label{sect:pheno}

To date, a fit to the actual polarized DIS data has only been performed using the large-$N_c$ limit of KPS evolution equation \cite{Adamiak:2021ppq}. This work includes 122 data points from various polarized DIS measurements with $x<0.1$ and $Q^2>m^2_{\text{charm}}$. The initial condition for the type-1 polarized dipole amplitude of each light quark flavor takes the Born-inspired form of\footnote{The large-$N_c$ KPS equation only contains the type-1 polarized dipole amplitude, $G(x^2_{10},zs)$ and its neighbor dipole amplitude. In an actual fit, however, the amplitudes for different quark flavors must be distinguished.}
\begin{align}
G^{(0)}_q(x^2_{10},zs) &= a_q\ln\frac{zs}{\Lambda^2} + b_q\ln\frac{1}{x^2_{10}\Lambda^2} + c_q\,,
\end{align}
where $a_q$, $b_q$ and $c_q$ are obtained from a fit to the data at moderate $x$.

This work sees a significant improvement in the fit in terms of $\chi^2$ per degree of freedom, which decreases from 1.07 in \cite{Sato:2016tuz} to 1.01 \cite{Adamiak:2021ppq}\footnote{Since the former performs the fit for all values of Bjorken $x$, it employs 2515 data points \cite{Sato:2016tuz}. This is much greater than 122 data points employed in the latter, which only includes small-$x$ data \cite{Adamiak:2021ppq}.}, if the evolution is set to begin at $x_0=0.1$. Setting $x_0$ too high leads to a poor fit, which is expected because the KPS evolution was derived for small $x$.  

Most importantly, the fit is capable to constraint the quark helicity's contribution to the proton spin\footnote{Note that no such conclusion can be drawn about the gluon helicity's contribution because the KPS equation separates the quark hPDF from the gluon's \cite{Kovchegov:2015pbl, Kovchegov:2018znm, Kovchegov:2017lsr}, in contrast to the KPSCTT equation in which they essentially couple to each other \cite{Cougoulic:2022gbk}.}. For instance, at scale $Q^2=10$ GeV$^2$, the quark helicity is expected to account for negative 10-20\% of the proton spin, as long as the starting point, $x_0$, of the KPS evolution is close to $0.1$, where the fit to the polarized DIS data is the best.

In light of the recent theoretical development, particularly the addition of the type-2 dipole, an updated fit is being performed with the large-$N_c\& N_f$ limit of the KPSCTT equation \cite{newJAM}. Due to the significant increase in the number of degrees of freedom, the fit requires additional data to sufficiently constraint the parameters. Hence, a new fit will employ not only the polarized DIS but also the polarized SIDIS data. Finally, a running coupling prescription is necessary to tame the growth of hPDFs in the KPSCTT equation until the SLA corrections are included.

\section{Conclusion}\label{sect:conclusion}

In this article, we outline our method to construct the small-$x$ helicity evolution equation in the CGC framework. Within the shockwave formalism, the evolution equation was derived, first by KPS involving only the type-1 dipole amplitude \cite{Kovchegov:2015pbl, Kovchegov:2018znm, Kovchegov:2017lsr}. More recently, the type-2 dipole amplitude is added, leading to the KPSCTT equation \cite{Cougoulic:2022gbk}. 

Both versions of the helicity evolution equation do not close in general, but instead form hierarchies of evolution equations that depend on increasingly complicated objects besides the $q\bar{q}$ dipoles. However, the equations become a closed system of linear integral equations in the large-$N_c$ and the large-$N_c\& N_f$ limits with the mean-field approximation. In each limit, the KPSCTT evolution still leads to a minor difference in the asymptotic behaviors of hPDFs and $g_1$ \cite{Cougoulic:2022gbk, newNcNf}, when compared to the BER helicity evolution \cite{Bartels:1996wc}. The cause of the disagreement is under investigation. 

A fit to the polarized DIS data has been performed successfully for the KPS evolution at large $N_c$ \cite{Adamiak:2021ppq}. An updated fit using the large-$N_c\& N_f$ KPSCTT evolution is in progress, employing the polarized DIS and SIDIS data. The new version of the fit will also include running coupling prescriptions to the evolution.

The SLA corrections have been calculated for the KPS evolution \cite{Kovchegov:2021lvz}, and the additional contributions required for the KPSCTT evolution are in progress \cite{SLAops}. The resulting equation will incorporate non-linear small-$x$ evolution, including saturation effects, into the formalism.

Finally, the framework can be modified to calculate the parton's orbital angular momentum at small $x$ \cite{Kovchegov:2019rrz}, whose update is also in progress to be consistent with the KPSCTT evolution \cite{BrandonOAM}. With all these theoretical results, together with future polarized small-$x$ measurements from the EIC \cite{Accardi:2012qut}, a clearer picture of the small-$x$ contribution to the proton spin puzzle will become available for all the terms in the Jaffe-Manohar sum rule \cite{JM}.

\section{Acknowledgments}

The author would like to thank Y. Kovchegov, D. Adamiak, J. Borden, F. Cougoulic, B. Manley, W. Melnitchouk, D. Pitonyak, N. Sato, M. Sievert and A. Tarasov for their works in various parts of the small-$x$ helicity research program. Furthermore, the author would like to thank the Cracow Epiphany Conference organizers for the opportunity to present the work.

The author is supported by the Academy of Finland, the Centre of Excellence in Quark Matter and project 346567, under the European Union’s Horizon 2020 research and innovation programme by the European Research Council (ERC, grant agreement No. ERC-2018-ADG-835105 YoctoLHC) and by the STRONG-2020 project (grant agreement No. 824093). The content of this article does not reflect the official opinion of the European Union and responsibility for the information and views expressed therein lies entirely with the authors.


\providecommand{\href}[2]{#2}\begingroup\raggedright\endgroup

\end{document}